\documentclass[12pt]{article}

\usepackage{amsmath,amssymb}
\usepackage{graphicx}
\usepackage{booktabs}
\usepackage{multirow}
\usepackage{float}
\usepackage{caption}
\usepackage{subcaption}
\usepackage{hyperref}
\usepackage{times}
\usepackage{url}
\usepackage{cite}
\usepackage{authblk}
\usepackage{adjustbox}
\usepackage{makecell}
\usepackage{geometry}
\title{Placenta Accreta Spectrum Detection using Multimodal Deep Learning}

\author[1*]{Sumaiya Ali}
\author[1*]{Areej Alhothali}
\author[2*]{Sameera Albasri}
\author[1]{Ohoud Alzamzami}
\author[3]{Ahmed Abduljabbar}
\author[3]{Muhammad Alwazzan}

\affil[1]{Department of Computer Science, Faculty of Computing and Information Technology,
King Abdulaziz University, Jeddah, Saudi Arabia}

\affil[2]{Faculty of Medicine,
King Abdulaziz University, Jeddah, Saudi Arabia}

\affil[3]{Department of Radiology, King Abdulaziz University Hospital\protect\\Jeddah, Saudi Arabia}

\affil[*]{sali0174@stu.kau.edu.sa, aalhothali@kau.edu.sa, salbasri@kau.edu.sa}

\date{}

\begin{document}
\maketitle

\begin{abstract}
Placenta Accreta Spectrum (PAS) is a life-threatening obstetric complication involving abnormal placental invasion into the uterine wall. Early and accurate prenatal diagnosis is essential to reduce maternal and neonatal risks. This study aimed to develop and validate a deep learning framework that enhances PAS detection by integrating multiple imaging modalities. A multimodal deep learning model was designed using an intermediate feature-level fusion architecture combining 3D Magnetic Resonance Imaging (MRI) and 2D Ultrasound (US) scans. Unimodal feature extractors, a 3D DenseNet121-Vision Transformer for MRI and a 2D ResNet50 for US, were selected after systematic comparative analysis. Curated datasets comprising 1,293 MRI and 1,143 US scans were used to train the unimodal models and paired samples of patient-matched MRI-US scans was isolated for multimodal model development and evaluation. On an independent test set, the multimodal fusion model achieved superior performance, with an accuracy of 92.5\% and an Area Under the Receiver Operating Characteristic Curve (AUC) of 0.927, outperforming the MRI-only (82.5\%, AUC 0.825) and US-only (87.5\%, AUC 0.879) models. Integrating MRI and US features provides complementary diagnostic information, demonstrating strong potential to enhance prenatal risk assessment and improve patient outcomes.
\end{abstract}


\section{Introduction}
The Placenta Accreta Spectrum (PAS) is a complex and life-threatening condition in which the placenta attaches too deeply to the uterine wall. The incidence of PAS has been steadily increasing, primarily due to the increase in cesarean section deliveries and the associated uterine scarring. It is estimated that the incidence has more than doubled over the past few decades~\cite{Cahill_2018}, with a 2022 study reporting a prevalence of 13.32 per 1000 births in Saudi Arabia~\cite{Basri_2022}. PAS poses significant risks during pregnancy and childbirth, including severe maternal hemorrhage, infection, the need for invasive procedures such as emergency hysterectomy, and increased maternal morbidity and mortality~\cite{Cahill_2018}. The underlying mechanisms of PAS are still not fully understood~\cite{Zhu_2022} and no reliable protein markers have been identified to date~\cite{YAŞAR_2022}.

Early and accurate detection of PAS is crucial for timely intervention and optimal management of affected pregnancies~\cite{Cahill_2018}. Traditional diagnostic methods involve a combination of clinical evaluation, biochemical markers, Ultrasound (US) imaging, and Magnetic Resonance Imaging (MRI)~\cite{Arakaza_2023}. However, these methods often require subjective interpretation and can have limitations in accurately predicting the extent and severity of placental invasion. The complexity of PAS requires a multidisciplinary approach; however, the integration of data and expertise from various fields to combine the results of different tests and imaging methods in order to provide a timely and accurate diagnosis remains a challenge~\cite{Morlando_2020, Wang_2023}. This can lead to delayed or inadequate management that increases maternal and fetal complications~\cite{Wang_2023}. Given these challenges, artificial intelligence (AI), especially deep learning, has become a promising objective way to improve the diagnosis and clinical decisions in PAS.

Deep learning models, particularly convolutional neural networks (CNN), have shown remarkable capabilities in learning intricate patterns and features from complex medical images including detection of PAS from different data sources~\cite{Pei_2023}. Nevertheless, limitations still exist where traditional machine learning and deep learning models often rely on a single modality such as US images or MRI to make predictions. This unimodal approach can result in the loss of relevant information that could otherwise improve diagnostic performance. New innovative deep learning models have been developed that use multiple modalities for detection and classification in various medical diagnostic tasks~\cite{Venugopalan_2021, Holste_2021, Yan_2021}. Research shows that multimodal learning can improve the performance of unimodal learning~\cite{Pei_2023}. Training deep learning models on extensive and heterogeneous datasets composed of both US and MRI images enables the development of robust and accurate algorithms for PAS detection~\cite{Wang_2023}. However, current research on PAS detection using multimodal deep learning frameworks that combine the complementary strengths of different imaging modalities remains limited.

Recent studies have mostly focused on single imaging modalities, particularly MRI, which is often considered superior to US when US findings are inconclusive~\cite{Romeo_2021}. Some studies~\cite{Romeo_2019, Leitch_2022} demonstrated that radiomic features from T2-weighted imaging (T2WI) sequences of MRI could predict PAS with high accuracy, while subsequent studies combined both radiomic and deep features~\cite{Shao_2021, Peng_2023}. Peng et al.~\cite{Peng_2023} presented a deep learning radiomics model using MRI that extracted radiomic features via transfer learning with MedicalNet, achieving 0.861 AUC and 75\% accuracy on an independent test set and 0.852 AUC on external validation. More recent approaches employed segmentation networks (e.g., nnU-Net) combined with deep classifiers, achieving Area Under the Receiver Operating Characteristic Curves (AUC)s above 0.86 and accuracy 84.3\% that outperformed radiologists~\cite{Wang_2023}. Xu et al.~\cite{Xu_2022} fused T1-weighted imaging (T1WI) and T2WI sequences of MRI features using a dual-path neural network, achieving 82.5\% accuracy, demonstrating the potential of multi-sequence MRI data integration. 

Compared to MRI, fewer studies have explored US-based detection, despite its accessibility and cost-effectiveness, due to operator dependence and lower sensitivity~\cite{Ricciardi_2020, Yang_2022}. A recent study~\cite{Young_2024} developed machine learning models using texture features from 2D US images, reporting stable cross-validation accuracies of 83.0\% and 88.7\% for classical and linear classifiers including the reported 92.3\% and 87.2\% test results. Furthermore, radiomic features and k-nearest neighbors classification approaches have achieved over 80\% accuracy~\cite{Ricciardi_2020}, while deep dictionary learning and placental thickness analysis using US have also shown promise~\cite{Yang_2022, Elmaraghy_2023}. Beyond imaging, several studies have investigated clinical and biochemical predictors of PAS. Shazly et al.~\cite{Shazly_2021} built predictive models for hemorrhage risk using multicenter clinical data, while YAŞAR et al.~\cite{YAŞAR_2022} applied machine learning to plasma proteins, identifying potential biomarkers.  

While these prior works have explored single-modality data, very few have examined PAS detection using multimodal approaches. Ye et al.~\cite{Ye_2022} combined MRI and clinical data using ensemble learning, but applied only decision-level (late) fusion, achieving improved performance (AUC 0.857) over unimodal models. Zhu et al.~\cite{Zhu_2022} combined MRI radiomics and clinical variables using nnU-Net segmentation and logistic regression, achieving AUC 0.849, outperforming radiologists (AUC 0.744). In other domains, multimodal deep learning has been applied successfully. Holste et al.~\cite{Holste_2021} showed that intermediate feature fusion of breast MRI and clinical data improved diagnostic accuracy (AUC 0.898); Yan et al.~\cite{Yan_2021} integrated pathological images with clinical features for breast cancer classification, achieving 92.9\% accuracy; and Venugopalan et al.~\cite{Venugopalan_2021} fused MRI, genetic data, and clinical tests for Alzheimer's disease diagnosis, with multimodal models consistently outperforming unimodal ones. 

Existing work demonstrates that deep learning, particularly CNN-based approaches, outperforms traditional machine learning in PAS detection. However, most studies remain modality-specific, rely on manual segmentation, and/or use limited datasets. While prior work has combined imaging with clinical data using late-fusion or ensemble methods~\cite{Ye_2022, Zhu_2022}, and multimodal deep learning has proven effective in other medical domains~\cite{Pei_2023}, the development and validation of an end-to-end, feature-level fusion model combining US and MRI imaging for PAS detection remains a critical, unaddressed research gap~\cite{Danaei_2025, Jittou_2025}. This gap motivates the approach presented in this work.

The main motivation behind this study is to increase the accuracy and reliability of PAS detection by taking advantage of deep learning and multimodal data to help healthcare practitioners make informed decisions and deliver appropriate prenatal care. Integrating data from various sources will enable a more detailed view of a patient's condition, enhance diagnostic accuracy, and potentially reduce false positives and negatives~\cite{Wang_2023}. Therefore, this study develops a multimodal deep learning model that uses an intermediate feature fusion method to identify PAS from a curated dataset of 1,293 MRI and 1,143 US images.

The contributions of this study include the development of a dual-modality dataset of PAS and control cases incorporating both MRI and US images; the development of a multimodal deep learning model trained on this dataset for automated PAS detection; a comprehensive evaluation of the model's performance using accuracy, precision, recall, and other relevant metrics; and a comparative analysis demonstrating the advantages of the proposed multimodal approach over corresponding unimodal models.

\section{Methods}

\subsection{Data Collection}

\begin{figure}[ht]
    \centering
    \includegraphics[width=0.85\textwidth]{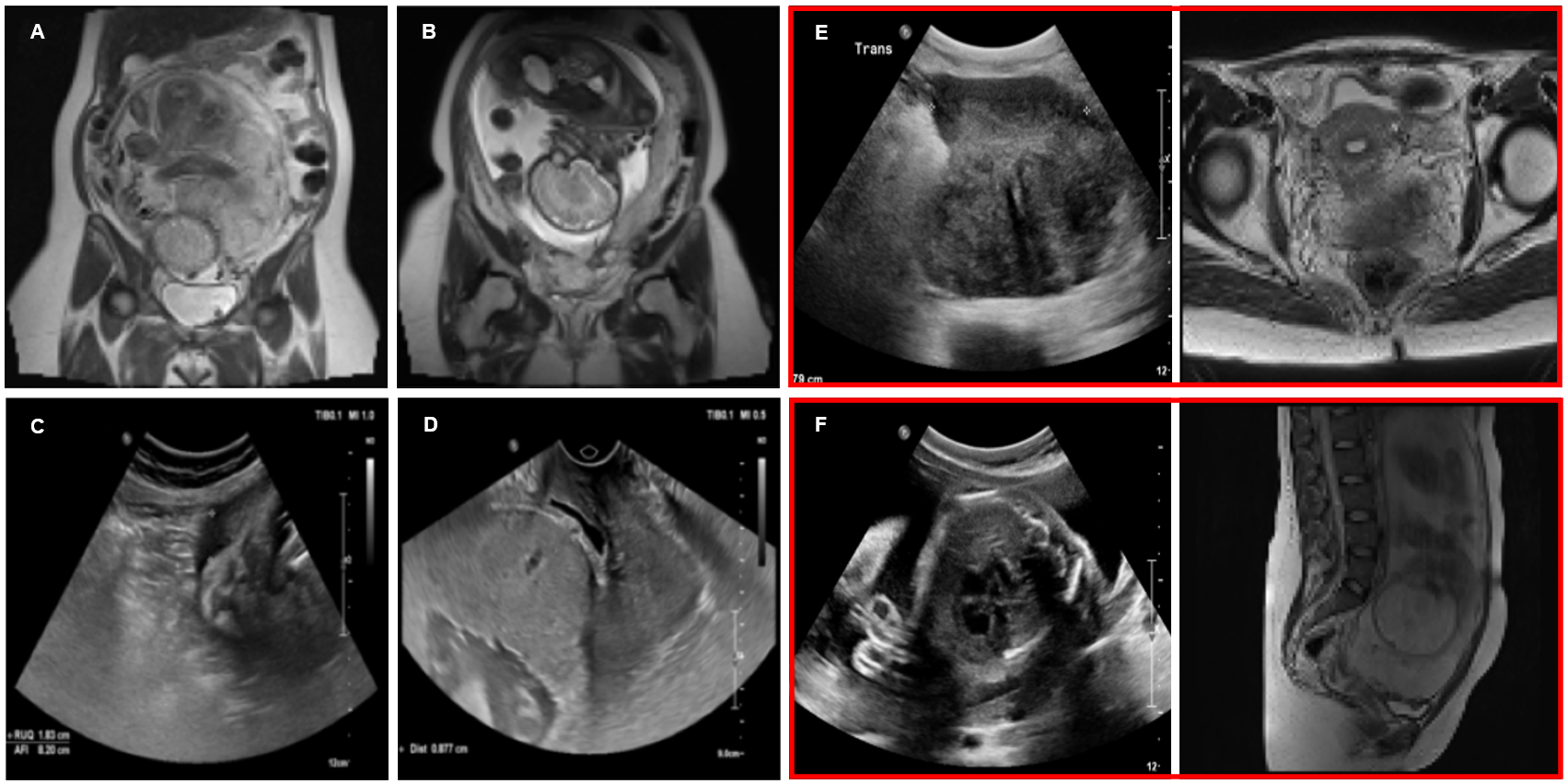} 
    \caption{Example of non-PAS MRI (A), PAS-positive MRI (B), non-PAS US (C), PAS-positive US (D), non-PAS multimodal pair (E), and PAS-positive multimodal pair (F).}
    \label{fig:data_example}
\end{figure}

A retrospective cohort of patients' data, collected in collaboration with the Department of Fetal Medicine at King Abdulaziz University Hospital, was used in this study. All patients' data were collected and used in accordance with institutional guidelines and ethical standards. Ethical approval was secured prior to the start of the study, and strict measures were taken to ensure patient confidentiality and compliance with data privacy regulations. A preliminary review of the health records in the hospital revealed a high proportion of patients who had been assessed with suspected PAS disorders. After applying rigorous inclusion criteria to select cases with confirmed diagnostic outcomes and sufficient image quality for computational analysis, the initial data comprised 1,293 T2WI MRI scans and 1,143 US images. Figure~\ref{fig:data_example} presents a few examples of the MRI and US collections, with the MRI displaying a single slice of the 3D scan. On MRI, PAS is commonly identified by intraplacental dark bands, interruption of the myometrial border and abnormal vascularity (Figure~\ref{fig:data_example}B), where as non-PAS cases show smooth uninterrupted utero-placental surface (Figure~\ref{fig:data_example}A). On US, PAS cases (Figure~\ref{fig:data_example}D) often present prominent irregular spaces of various sizes in placenta known as placental lacunae, thinning of myometrial thickness and loss of retroplacental clear space, while non-PAS image retains a uniform placental appearance and intact retroplacental zone (Figure~\ref{fig:data_example}C). The multimodal image pairs (Figure~\ref{fig:data_example}E-F) further highlight the complementary diagnostic information provided by MRI and US for distinguishing PAS-positive and non-PAS cases~\cite{Romeo_2021, Ali_2025}. The data collection was intended to represent a wide range of patients, both confirmed PAS and Normal (non-PAS), to develop robust and generalizable deep learning models that can perform accurate diagnosis in a clinical environment.

\subsection{Data Preparation and Preprocessing}

After the initial data collection, three separate datasets were created: a unimodal MRI dataset (1,133 scans), a unimodal US dataset (983 scans), and a paired multimodal dataset (160 pairs) in which each sample pair consists of an MRI scan and an US scan of the same patient. The unimodal datasets were kept larger than the multimodal dataset. Large unimodal datasets are essential for training deep feature extractors capable of learning the complex patterns associated with PAS from a single imaging modality. Then a smaller, high-quality paired dataset can be used for the final, more nuanced task of data fusion. This hierarchical approach allows the study to first maximize the knowledge extracted from large volumes of modality-specific data before learning the relationships between the modalities in a more focused, patient-matched data. A standardized, modality-specific preprocessing pipeline was implemented for all images to ensure data consistency, normalized intensity distributions, and an appropriate format for input to deep learning models.

\subsubsection{MRI Preprocessing}

\begin{figure}[htp!]
    \centering
    \includegraphics[width=\textwidth]{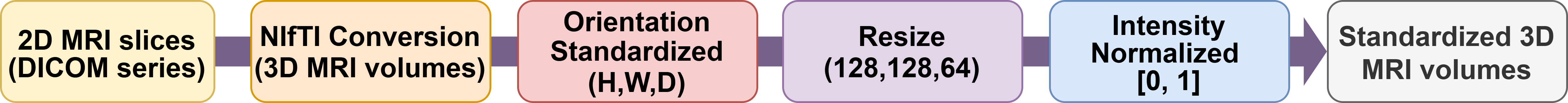} 
    \caption{Preprocessing pipeline for standardizing MRI scans.~\cite{Ali_2025}}
    \label{fig:mri_preprocessing}
\end{figure}

Raw 3D MRI was stored in the Digital Imaging and Communications in Medicine (DICOM) format in the form of 2D slices. These DICOM series were converted to the Neuroimaging Informatics Technology Initiative (NIfTI) format, which combined the individual 2D slices into a single 3D volumetric file suitable for 3D CNNs. To eliminate variability arising from different scan orientations, the axes of each 3D volume were rearranged to a standard (height, width, depth) order. All volumes were resized to a fixed dimension of $128\times128\times64$ voxels using cubic interpolation, with padding applied as needed to maintain original aspect ratio. Finally, the maximum voxel intensity for each scan was identified and the intensities were normalized to a floating-point value in the range 0 to 1 using min-max scaling. This spatial standardization is a prerequisite for batch processing in deep learning models with fixed-size input layers (see Figure \ref{fig:mri_preprocessing})~\cite{Ali_2025}.

\subsubsection{Ultrasound Preprocessing}

\begin{figure}[htp!]
    \centering
    \includegraphics[width=\textwidth]{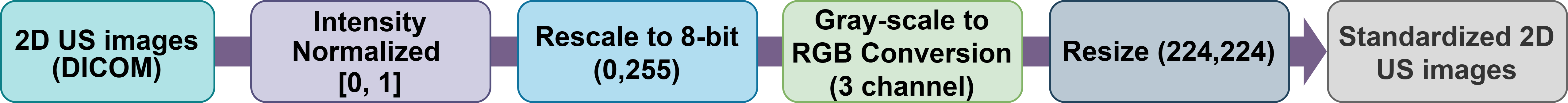} 
    \caption{Preprocessing pipeline for standardizing US images.}
    \label{fig:us_preprocessing}
\end{figure}

The 2D US images were processed through a different pipeline (see Figure \ref{fig:us_preprocessing}). Pixel intensities were first normalized using a min-max scaler and rescaled to 8-bit integers in the range 0 to 255. The single-channel grayscale images were then converted to a three-channel RGB format and resized to a resolution of $224\times224$ pixels, which is a standard input format for many of the pretrained CNN models evaluated in this study.

\subsection{Data Distribution and Augmentation}

The splitting of these datasets was performed using stratified sampling to ensure that the original class distribution was preserved across the training, validation, and test sets, thus preventing sampling bias. Different split ratios were used to accommodate the varying sizes and specific analytical purposes of the datasets. This stage also included data augmentation and modeling strategies to address the class imbalance evident in the datasets,  such as oversampling and class weighting during model training, aimed at preventing model bias towards the majority class. Table \ref{tab:dataset_summary} details the distribution of PAS (labeled 1) and Normal (labeled 0) cases across the training, validation, and test sets for the unimodal MRI, unimodal US, and paired multimodal datasets. 

\subsubsection{Unimodal MRI Dataset}

In the case of the unimodal MRI analysis, 1,133 distinct 3D MRI T2WI scans were chosen. There was a severe imbalance of classes in this dataset, as 853 scans were of the type Normal and 280 scans were PAS. The dataset was split into a training set (70\%; n = 793), a validation set (10\%; n = 113), and a hold-out test set (20\%; n = 227). 

\subsubsection{Unimodal US Dataset}

The unimodal US dataset was developed using a sample of 983 distinct 2D US scans. There was also a class imbalance in this dataset, as there were 676 scans of Normals and 307 scans of PAS. The data was divided in a training set (n = 687) and a validation set (n = 99) and a test set (n = 197), following the same 70:10:20 split ratio.

\subsubsection{Multimodal Dataset}

A smaller sample size of 160 paired samples was isolated from the larger MRI and US sample collections to develop and test the multimodal fusion model. Each sample in this data set consists of an MRI and a US scan of the same patient, so the anatomy of the two modalities is directly matched. This patient-level matching is essential for the model to acquire the complementary diagnostic information of each modality. The distribution of classes consisted of 100 Normal pairs and 60 PAS pairs. A distinct split ratio was adopted for this smaller dataset. The data was split into 60\% for training (n = 96), 15\% for validation (n = 24), and 25\% for testing (n = 40). A higher percentage of data was selected for the test set to focus on a more statistically sound estimate of the overall generalization performance of the final model, which is critical when testing models on smaller datasets.

\begin{table}[h!]
\centering
\caption{Dataset distribution across training, validation, and test sets.}
\label{tab:dataset_summary}
\begin{tabular}{l l c c c c}
\hline
\textbf{Dataset} & \textbf{Class} & \textbf{Training Set} & \textbf{Validation Set} & \textbf{Test Set} & \textbf{Total} \\
\hline
\textbf{Unimodal MRI} & Normal (0) & 597 & 85 & 171 & 853 \\
 & PAS (1) & 196 & 28 & 56 & 280 \\
 & \textbf{Total} & \textbf{793} & \textbf{113} & \textbf{227} & \textbf{1133} \\
\hline
\textbf{Unimodal US} & Normal (0) & 473 & 68 & 135 & 676 \\
 & PAS (1) & 214 & 31 & 62 & 307 \\
 & \textbf{Total} & \textbf{687} & \textbf{99} & \textbf{197} & \textbf{983} \\
\hline
\textbf{Multimodal} & Normal (0) & 60 & 15 & 25 & 100 \\
 & PAS (1) & 36 & 9 & 15 & 60 \\
 & \textbf{Total} & \textbf{96} & \textbf{24} & \textbf{40} & \textbf{160} \\
\hline
\end{tabular}
\end{table}

\subsubsection{Data Augmentation}

To minimize the risk of overfitting and deal with imbalances in classes, a set of data augmentation methods was used in the training sets of the unimodal datasets. In the case of the MRI dataset, data augmentation and oversampling of the minority class (PAS) were used to deal with the class imbalance in the training set. PAS samples were synthetically oversampled from 196 to 597 samples to obtain a final balanced training set of 1,194 volumes. The geometric transformations that were performed on the 3D MRI volumes were random flips along the height and width axis, random rotation (90\textdegree, 180\textdegree, 270\textdegree) and random zoom (factors 1.1-1.3).

In the case of the US data training set, augmentations were random horizontal flips and random rotations in a continuous range of ±10 degrees. Class imbalance was solved algorithmically using class weights during loss computation. This method puts a heavier penalty on misclassification of samples belonging to the minority class (PAS), which causes the model to better learn its distinguishing characteristics.

\subsection{Deep Learning Model Architectures}

In order to determine whether fusion can enhance the prediction of PAS, unimodal baseline models are initially developed. It involved a systematic search through different state-of-the-art deep learning architectures to find the most successful feature extractors of each modality. Each imaging modality was evaluated on these architectures. The primary strategy consisted of fine-tuning models that were pretrained on large scale public data to take advantage of their powerful, existing feature extraction properties, therefore enhancing performance and reducing the amount of training data needed. The resulting selected unimodal models in this process guided the development of a novel multimodal fusion model.

\subsubsection{MRI Feature Extraction Model}

In the case of the volumetric MRI data, six 3D architectures were explored. These included pure CNN 3D ResNet18~\cite{He_2016}, 3D DenseNet121~\cite{Huang_2017} and 3D EfficientNet-B0~\cite{tan_2019}; hybrid models like 3D DenseNet121 with 3D Vision Transformer (ViT) and 3D ResNet18 with 3D Swin transformer~\cite{Ahmed_2025}; and a pure transformer model, the 3D Swin transformer~\cite{liu2021swin}. To reduce overfitting, regularization techniques such as dropout layers were incorporated~\cite{Ali_2025}. Various dropout rates in the range of (10\%-50\%) were examined for the 3D DenseNet121-ViT and 3D ResNet18. The optimal dropout rates were found to be 50\% for 3D DenseNet121-ViT and 10\% for 3D ResNet18. The 3D ResNet18 model was pretrained using the MedicalNet~\cite{chen_2019, Ali_2025}.

\begin{figure}[htp!]
    \centering
    \includegraphics[width=\textwidth]{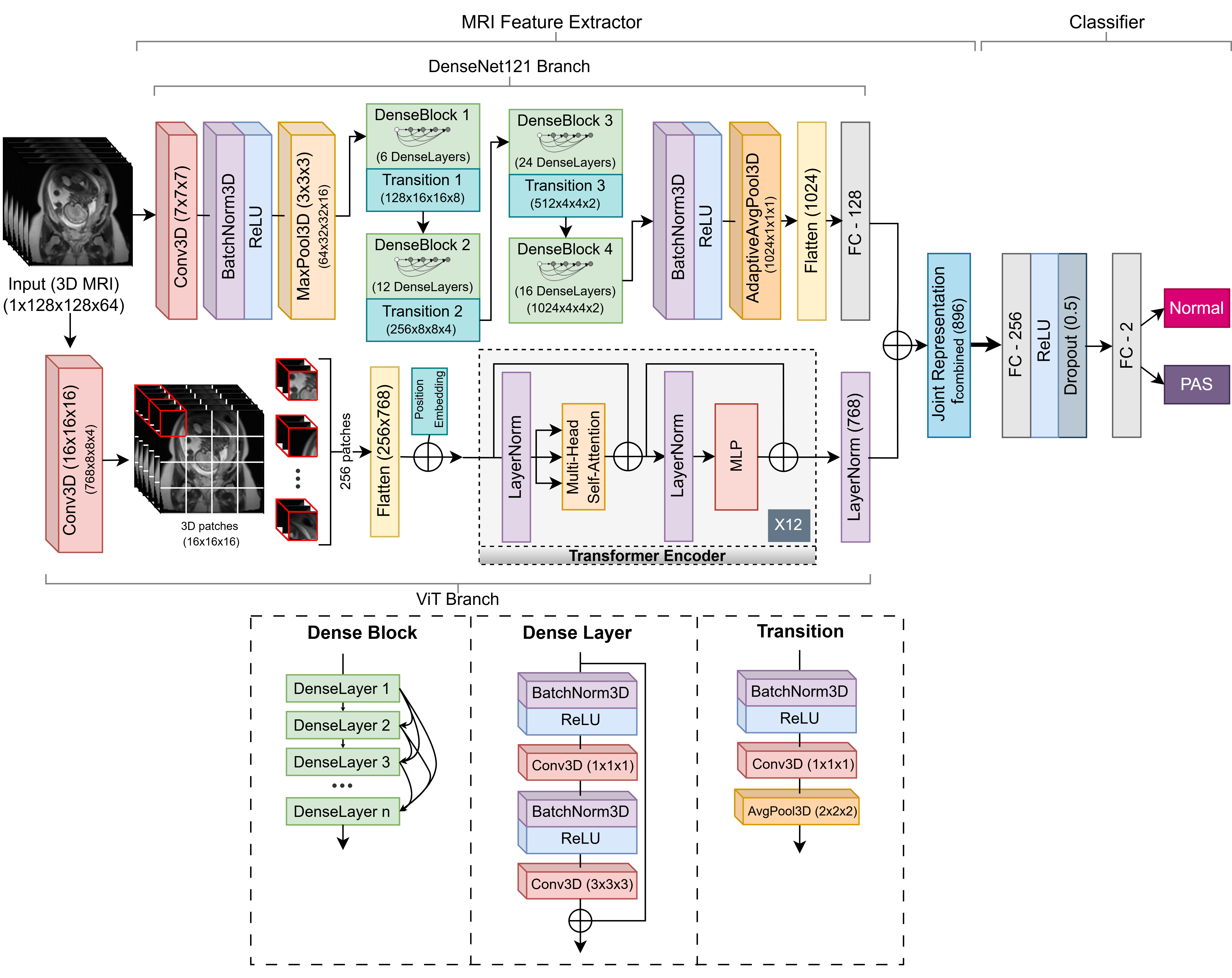} 
    \caption{The selected MRI model architecture of 3D DenseNet121-ViT.~\cite{Ali_2025}}
    \label{fig:densenetvit_model}
\end{figure}

Following the comparative analysis detailed in Results section, the hybrid 3D DenseNet121-ViT architecture was selected as the feature extraction backbone for the MRI data. The simplified structure of the selected MRI model is shown in Figure \ref{fig:densenetvit_model}. The hybrid 3D DenseNet121-ViT model has two parallel modules of DenseNet121 and ViT that takes MRI volumes as input before fusing their extracted features for joint processing. The 3D MRI inputs are denoted as $\mathbf{X} \in \mathbb{R}^{C \times H \times W \times D}$, where H is height = 128, W is weight = 128, D is depth = 64 and C is channel dimension = 1 (grayscale).

\textbf{DenseNet121 Branch}: The DenseNet121 branch extracts high-level local features through densely connected convolutional layers. The initial feature extraction stage applies a 3D convolutional layer (Conv3D) to capture low-level spatial features while performing downsampling. For the input volume $X$, the Conv3D operation is defined as in equation (\ref{eq:3dconv}), where $X_{conv}{(i,j,l,f)}$ represents the activation at spatial coordinates $(i, j, l)$ for the $f^{th}$ feature map, $\sigma$ denotes the Rectified Linear Unit (ReLU) activation function, $s$ is stride, $W(m,n,p,c,f)$ denotes the learnable weight of the filter at the local kernel position $(m, n, p)$ for the $c^{th}$ input channel and $f^{th}$ output feature map and $b_f$ is the bias term for the $f^{th}$ feature map. The kernel $W$ has a spatial size of $K \times K \times K$ with $K=7$. By utilizing a stride of $s=2$, the layer reduces the spatial dimensions of the input $X$ by half $(\frac{H}{2} \times \frac{W}{2} \times \frac{D}{2})$, outputting 64 feature maps. The batch normalization (BatchNorm3D) and ReLU activation function normalizes the outputs and applies non-linearity. 

{\footnotesize
\begin{equation} \label{eq:3dconv}
X_{conv}{(i,j,l,f)} = \sigma \left( \sum_{c=1}^{C} \sum_{m=0}^{K-1} \sum_{n=0}^{K-1} \sum_{p=0}^{K-1} X({i \cdot s + m, \,\, j \cdot s + n, \,\, l \cdot s + p, \,\, c}) \cdot W({m,n,p,c,f}) + b_f \right)
\end{equation}}

The max-pooling layer (MaxPool3D) further reduces spatial dimensions by half with equation (\ref{eq:maxpool}), where, $X_{max}{(i,j,l,c)}$ is the downsampled output value for the $c^{th}$ channel at coordinates $(i, j, l)$, the indices $m, n, p$ represent the spatial coordinates within the window, while $k=3$ and $s=2$ denote the pooling kernel size and stride, respectively. The max operation chooses the maximum value within the defined window. This process results in a final output of volume $(\frac{H}{4} \times \frac{W}{4} \times \frac{D}{4})$, which is $(32 \times 32 \times 16)$ for each of the $64$ feature maps, before being passed into the dense blocks.

\begin{equation} \label{eq:maxpool}
X_{max}{(i,j,l,c)} = \max_{0 \leq m, n, p < k} \{ X_{conv}{(i \cdot s + m, j \cdot s + n, l \cdot s + p,c)} \}
\end{equation}

The features $X_{max}$ are fed into a series of four dense blocks (DenseBlock 1 to 4) each having $L$ = 6, 12, 24 and 16 dense layers, respectively. The dense blocks are densely connected layers where each $l^{th}$ layer takes as input all previous layers outputs using concatenation. This recursive pattern for each dense layer is expressed in equation (\ref{eq:dense_conn}), where $[x_0, x_1, \dots, x_{\ell-1}]$ denotes the concatenation of feature maps, with $x_0 = X_{max}$ and $\mathcal{F}_\ell(\cdot)$ is the function of the $\ell^{th}$ layer consisting of BatchNorm3D, ReLU activation and Conv3D. The convolution operations in each dense layer include a $1 \times 1 \times 1$  Conv3D to reduce the number of input feature maps and a $3 \times 3 \times 3$  Conv3D to extract spatial features. Each layer produces $k$ feature maps (where $k=32$) and the $x_\ell$ have a channel dimension of $k_0 + (\ell \times k)$, where $k_0$ is the number of input channels. All use stride 1, meaning no reduction of spatial dimensions in the blocks. The dense block outputs, $X_{dense}$, are the aggregation of all generated features in $L$ layers of that block (equation (\ref{eq:x_dense})).

\begin{equation} \label{eq:dense_conn}
x_\ell = \mathcal{F}_\ell([x_0, x_1, \dots, x_{\ell-1}])
\end{equation}

\begin{equation} \label{eq:x_dense}
X_{dense} = [x_0, x_1, x_2, \dots, x_L]
\end{equation}

Between dense blocks, transition layers (Transition 1 to 3) perform downsampling and feature compression using 3D average pooling (AvgPool3D) as in equation (\ref{eq:transition}) that computes the mean of all values in the given window. Here, $\mathcal{G}(\cdot)$ represents the composite function of the transition layer on the output of the preceding dense block, $X_{dense}$. This consists of BatchNorm3D, ReLU activation and $1 \times 1 \times 1$ Conv3D with stride 1 used for feature compression. The $X_{trans}$ at position $(i, j, l)$ for channel $c$ is the output value after average pooling with $k=2$ and $s=2$ as kernel size and stride, while $m, n, p$ are the spatial coordinates. The Conv3D layer reduces the channel dimension by half while AvgPool3D layer reduces spatial dimension by half.

\begin{equation} \label{eq:transition} 
X_{trans}(i, j, l, c) = \frac{1}{k^3} \sum_{m=0}^{k-1} \sum_{n=0}^{k-1} \sum_{p=0}^{k-1} \mathcal{G}(X_{dense})(i \cdot s + m,j \cdot s + n,l \cdot s + p,c) \end{equation}

The output of the final dense block (DenseBlock 4) is a feature map $\hat{X} \in \mathbb{R}^{C \times H \times W \times D}$, where $H=4, W=4, D=2$ and $C=1024$, undergoes adaptive average pooling (AdaptiveAvgPool3D) as in equation (\ref{eq:avgpool3d}) that computes the average across the entire dimension of each feature map. 

\begin{equation} \label{eq:avgpool3d} 
v_c = \frac{1}{H \cdot W \cdot D} \sum_{i=1}^{H} \sum_{j=1}^{W} \sum_{l=1}^{D} \hat{X}{(i,j,l,c)}
\end{equation}

The output of AdaptiveAvgPool3D ($v_c$) then flattened to produce a feature vector $f_{flatten} \in \mathbb{R}^{1024}$. A fully connected (FC) layer then compresses this into a 128-dimensional local feature embedding with equation (\ref{eq:dense_fc}), where, $W_{fc}$ and $b_{fc}$ represent the weights and biases of the FC layer.

\begin{equation} \label{eq:dense_fc} 
f_{dense} = \sigma(W_{fc} \cdot f_{flatten} + b_{fc}), \quad
f_{dense} \in \mathbb{R}^{128}
\end{equation}

\textbf{ViT Branch}: The ViT branch takes the 3D MRI and processes it as a sequence of 3D patches. The input volume $X$ is divided into $N$ non-overlapping cubic patches of size $P \times P \times P$ using Conv3D layer with kernel $16 \times 16 \times 16$ and stride 16. The total number of patches is calculated as in equation (\ref{eq:patch_count}). A total of 256 patches are generated each of size $16 \times 16 \times 16$. Each patch is linearly projected into a 768-dimensional embedding (a token). Flatten layer flattens them into a sequence of tokens (each representing one patch). This operation is defined in equation (\ref{eq:patch_emb}), where $\mathbf{p}_i \in \mathbb{R}^{P \times P \times P}$ denotes the $i$-th 3D patch, 
$\mathbf{W}_p$ and $\mathbf{b}_p$ are learnable parameters, and $d = 768$ is the embedding dimension. To preserve 3D spatial context in the flattening process, a 3D positional embedding ($\mathbf{e}_i^{pos}$) is added to each token which tells the transformer where each patch came from in the 3D volume (equation (\ref{eq:pos_emb})).

\begin{equation} \label{eq:patch_count} 
N = \frac{H \cdot W \cdot D}{P^3} = \frac{128}{16} \times \frac{128}{16} \times \frac{64}{16} = 8 \times 8 \times 4 = 256 
\end{equation}
\begin{equation} \label{eq:patch_emb} 
\mathbf{e}_i = \mathbf{W}_p \cdot \text{Flatten}(\mathbf{p}_i) + \mathbf{b}_p,
\quad \mathbf{e}_i \in \mathbb{R}^{d}
\end{equation}
\begin{equation} \label{eq:pos_emb}
\tilde{\mathbf{e}}_i = \mathbf{e}_i + \mathbf{e}_i^{pos}
\end{equation}

The resulting embedded tokens sequence, $\{\tilde{\mathbf{e}}_1, \tilde{\mathbf{e}}_2, \ldots, \tilde{\mathbf{e}}_N\}$ is then passed to the transformer encoder for global contextual modeling. There are 12 transformer encoder blocks stacked sequentially. Each block normalizes the input token features with LayerNorm (LN) before the Multi-Head Self-Attention (MHSA). The MHSA computes attention among all tokens (patches) to capture spatial relationships between them. The self-attention operations is defined in equation (\ref{eq:self_att})~\cite{Aloraini_2023}, where Q, K and V are query, key and value matrices computed from input $\mathbf{E} \in \mathbb{R}^{N \times d}$, which is the matrix formed by stacking the token embeddings, and $\mathbf{W}_Q$, $\mathbf{W}_K$, and $\mathbf{W}_V$ are learnable projection matrices.
\begin{equation} \label{eq:self_att}
\mathbf{Q} = \mathbf{E}\mathbf{W}_Q, \quad
\mathbf{K} = \mathbf{E}\mathbf{W}_K, \quad
\mathbf{V} = \mathbf{E}\mathbf{W}_V, \quad
\text{MHSA}(\mathbf{Q}, \mathbf{K}, \mathbf{V})
=
\text{Softmax}\!\left(
\frac{\mathbf{Q}\mathbf{K}^\top}{\sqrt{d}}
\right)\mathbf{V}
\end{equation}

The output of the MHSA layer is then combined with the input tokens via a residual connection and LN (equation (\ref{eq:mhsa})). Each encoder block further applies a position-wise Multi-Layer Perceptron (MLP) resulting in the final output of the block (equation (\ref{eq:mlp})). The MLP block applies two FC layers on each token independently and Gaussian Error Linear Unit activation function that adds non-linear transformation capacity. The output ${E}_{\text{out}}$ after 12 Transformer Encoder blocks is normalized with LN producing the 768-dimensional ViT feature vector representation ($f_{vit} \in \mathbb{R}^{768}$).

\begin{equation} \label{eq:mhsa}
\mathbf{E}' = \text{LN}\!\left(\mathbf{E} + \text{Attention}(\mathbf{Q}, \mathbf{K}, \mathbf{V})\right)
\end{equation}
\begin{equation} \label{eq:mlp}
\mathbf{E}_{\text{out}} = \text{LN}\!\left(\mathbf{E}' + \text{MLP}(\mathbf{E}') \right)
\end{equation}

With both local features ($f_{dense}$) and global features ($f_{vit}$) extracted from the two branches, they are concatenated into a joint hybrid representation $f_{combined} \in \mathbb{R}^{896}$ (equation (\ref{eq:fusion})). The combined vector is passed through a classification head consisting of an FC layer (256 neurons), ReLU activation, and a Dropout layer ($p=0.5$). Final FC layer maps the 256-dimensional feature to 2 output neurons and the final prediction $\hat{y}$ for the class (PAS or Normal) is obtained via a softmax activation as in equation (\ref{eq:mri_softmax}).
\begin{equation} \label{eq:fusion}
f_{combined} = [f_{dense} \oplus f_{vit}]
\end{equation}
\begin{equation} \label{eq:mri_softmax}
\hat{y} = \text{Softmax}(\text{FC}_{896 \to 256 \to 2}(f_{combined}))
\end{equation}

\subsubsection{Ultrasound Feature Extraction Model}

\begin{figure}[htp!]
    \centering
    \includegraphics[width=\textwidth]{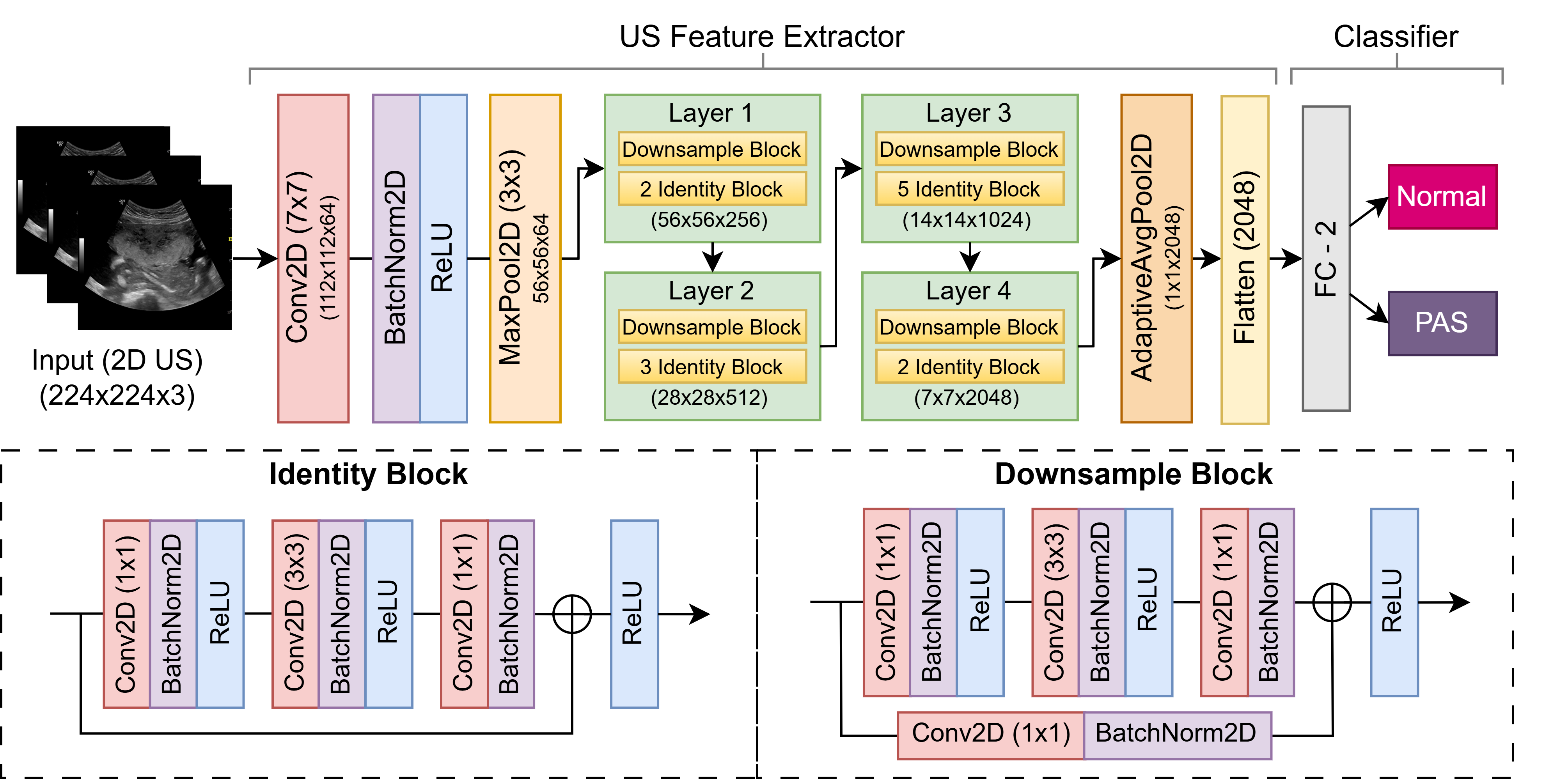} 
    \caption{The selected US model architecture of 2D ResNet50.}
    \label{fig:resnet50_model}
\end{figure}

For the 2D US data, a set of well-established 2D CNNs were trained and fine-tuned for the classification task. These included DenseNet121~\cite{Huang_2017} \& ResNet50~\cite{He_2016} pretrained on the RadImageNet~\cite{Mei_2022} database and ResNet18~\cite{He_2016} and EfficientNet-B0~\cite{tan_2019} pretrained on the ImageNet~\cite{deng_2009} database. ResNet50 architecture was selected as the feature extraction backbone the US data after careful analysis of the results. The simplified structure of the selected US model architecture is shown in Figure \ref{fig:resnet50_model}.

The 2D US image inputs, denoted as $\mathbf{I} \in \mathbb{R}^{H \times W \times C}$, have spatial dimensions of $H \times W = 224 \times 224$ with channel dimension $C$ = 3 for RGB. The input $\mathbf{I}$ is passed through the first layers of Conv2D and MaxPool2D layers. The Conv2D layer operation is represented by equation (\ref{eq:2dconv}), where $C=3$, $W$ is the learnable weight, kernel size $K=7$, $b_k$ is the bias term for the $k$-th filter, $s$ is the stride and $(m, n)$ are the local coordinates within the kernel. The Conv2D layer uses a $7\times7$ kernel and stride 2 that reduce the spatial dimensions by half. This layer outputs 64 feature maps. Batch normalization (BatchNorm2D) and ReLU activation function ($\sigma$) normalizes the outputs and applies non-linearity. The MaxPool2D layer further reduces spatial dimensions by half with kernel $3 \times 3$ and stride 2 resulting in the final output of ($\frac{H}{4}$, $\frac{W}{4}$, 64) (equation (\ref{eq:2dmax})).

\begin{equation} \label{eq:2dconv}
I_{conv}{(i,j,k)} = \sigma \left(  \sum_{c=1}^{C} \sum_{m=0}^{K-1} \sum_{n=0}^{K-1} I{(i \cdot s + m,j \cdot s + n, c)} \cdot W{(m,n,c,k)} + b_k \right)
\end{equation}

\begin{equation} \label{eq:2dmax}
I_{max}{(i,j,k)} = \max_{0 \leq m,n < K} \{ I_{conv}{(i \cdot s + m, j \cdot s + n, k)} \}
\end{equation}

There are four residual layers, each consisting of downsampling blocks and identity blocks. Layer 1 to 4 each has one downsampling block and 2, 3, 5 and 3 identity blocks respectively. The blocks have residual connection that adds the input to the block’s output. The operations in the downsampling and identity block are presented in equation (\ref{eq:downsample}) and equation (\ref{eq:identity}), respectively, where $\mathbf{x}$ is the input to the layers, $\mathbf{x'}$ is the outout after downsampling block, $I_{res}$ is the residual layer output and  $F(\mathbf{\cdot})$ represents the series of functions consisting three Conv2D layers with kernels $(1\times1)$, $(3\times3)$ and $(1\times1)$. Every Conv2D layer have BatchNorm2D and ReLU function following them. The $h(\mathbf{\cdot})$ in equation (\ref{eq:downsample}) refers to the extra Conv2D layer with kernel $(1\times1)$ in the residual connection that has strides $>1$ in order to reduces spatial size and increase channels. While identity blocks preserves size and channels and only learns features. 

\begin{equation} \label{eq:downsample}
\mathbf{x'} = \sigma(F(\mathbf{x}) + h(\mathbf{x}))
\end{equation}
\begin{equation} \label{eq:identity}
I_{res} = \sigma(F(\mathbf{x'}) + \mathbf{x'})
\end{equation}

The output $I_{res}$ from the last residual layer $(7\times7\times2048)$ passes to an adaptive average pooling (AdaptiveAvgPool2D) layer (equation (\ref{eq:avgpool})) that reduces the $7\times7$ feature maps to $1\times1$ by averaging each channel to single value ($z$). A FC layer maps the resulting flattened 2048-dimensional feature vector to 2 output classes (PAS and Normal) with softmax to produce the probabilities (equation (\ref{eq:softmax})).

\begin{equation} \label{eq:avgpool} 
z_c = \frac{1}{H \cdot W} \sum_{i=1}^{H} \sum_{j=1}^{W}I_{res}{(i,j,c)}
\end{equation}
\begin{equation} \label{eq:softmax}
\hat{z} = \text{Softmax}(\text{FC}_{2048 \to 2}(z))
\end{equation}

\subsubsection{Multimodal Intermediate Fusion Architecture}

Building upon the unimodal feature extractors, an intermediate fusion architecture (also known as feature-level fusion) was developed to integrate the information from both MRI and US modalities. This approach was chosen over early (input-level) or late (decision-level) fusion as it provides a balance between feature diversity and model complexity. The architecture is designed to identify and learn from the complementary relationships between the high-level abstract features extracted from each modality. This process computationally mirrors the diagnostic process of a multidisciplinary clinical team, where the findings of a sonography expert are combined with the MRI findings of a radiologist to arrive at a unified diagnosis. The model architecture was designed to replicate this combination of abstract clinical signs. 

\begin{figure}[htp!]
    \centering
    \includegraphics[width=0.75\textwidth]{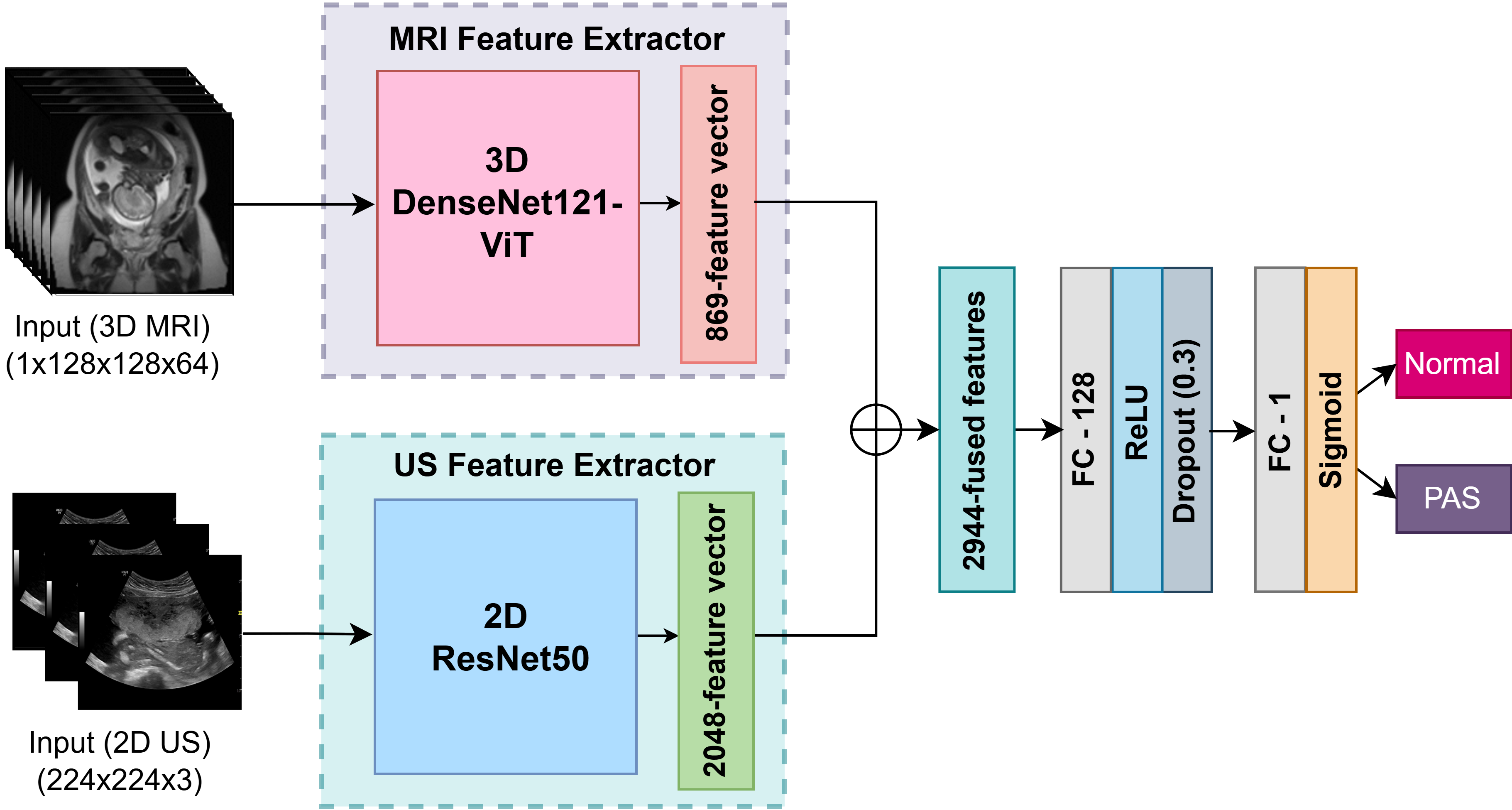} 
    \caption{The proposed multimodal architecture.}
    \label{fig:multimodal_model}
\end{figure}

The overall architecture of the proposed multimodal intermediate fusion model is illustrated in Figure~\ref{fig:multimodal_model}.
The multimodal fusion model was constructed as follows:
\begin{enumerate}
\item \textbf{Feature Extraction:} The fine-tuned unimodal MRI and US models served as the parallel feature extraction branches. As shown in Figures~\ref{fig:densenetvit_model} and~\ref{fig:resnet50_model}, all layers in the MRI and US Feature Extractor blocks were retained, while the unimodal classifiers were removed. Their weights were left unfrozen to allow for end-to-end fine-tuning where the unimodal branches do further learning of features that are complementary to each other.
\item \textbf{Feature Vector Generation:} For a given patient's paired MRI and US scans, each scan was passed through its respective feature extractor to generate high-dimensional feature vectors that numerically represent the most important diagnostic information from each imaging modality. The MRI feature extractor generated 869-feature vector ($f_{mri}$) and the US generated 2048-feature vector ($f_{us}$).
\item \textbf{Feature Fusion:} The feature vector extracted from the MRI scan and the feature vector from the corresponding US scan are concatenated to form a single, unified feature vector: $f_{fused} = f_{mri}+f_{us}$. This joint representation contains a rich, complementary blend of information from both modalities.
\item \textbf{Classification:} The fused vector is then passed to an MLP which consisted of two FC dense layers. It compresses the fused features, learns complex non-linear relationships between the MRI \& US features, regularizes, and outputs a probability for classification. The fusion model uses the ReLU activation function, followed by a dropout layer with a rate of 30\% for regularization. The final output layer is a single-neuron FC layer with a sigmoid activation function, producing the binary prediction $Y$ for the presence or absence of PAS: 
\begin{equation} \label{eq:multi_clas}
Y = \text{Sigmoid}(\text{FC}_{2944 \to 128 \to 1}(f_{fused}))
\end{equation}
\end{enumerate}

\subsection{Experimental Setup and Training}

All experiments were performed in the PyTorch and MONAI~\cite{cardoso_2022} framework. The data were divided into training, validation and testing sets with stratified sampling to make sure that the data of one patient was not included in more than one set.

\subsubsection{Training Hyperparameter}

The training protocols and hyperparameter settings such as batch size, weight decay, and layer freezing strategies were systematically varied to identify the optimal configuration for each modality. The Adam optimization algorithm was used consistently across all experiments for its adaptive Learning Rate (LR) capabilities and robust performance in deep learning tasks. A default initial LR of $1 \times 10^{-4}$ was used for all models. A 'ReduceLROnPlateau' LR scheduler was utilized to adjust the LR by reducing it when a monitored metric, such as validation loss, stops improving, facilitating a more stable convergence to an optimal solution. Cross-Entropy Loss was chosen as the loss function for the unimodal models and Binary Cross-Entropy loss was employed for the multimodal model. A regularized version of the loss function with a label smoothing factor of 0.1 was used for the US model only for performance improvement due to higher noise observed in US imaging compared to MRI. Dropout was applied before the final FC layer with rate 0.5 for the MRI model and 0.3 for the multimodal model to mitigate overfitting. The optimal batch size for the given data sizes was found to be 8. Models were trained for an extensive number of epochs (ranging from 50 to 200) to ensure full convergence. The weights of the model were restored from the epoch with the highest accuracy on the validation set for final evaluation on the test set. The optimized hyperparameters used in the proposed approach are presented in Table \ref{tab:hyperparameters}. For the ViT in the hybrid 3D DenseNet121-ViT architecture, key hyperparameters included a patch size of 16x16x16, an embedding dimension of 768, 12 attention heads and transformer layers. 

\begin{table}[htbp]
\centering
\caption{Hyperparameters and their optimized values used in the proposed approach.}
\label{tab:hyperparameters}
\begin{tabular}{llll}
\hline
\textbf{Hyperparameters}        & \textbf{MRI model} & \textbf{US model} & \textbf{Multimodal model}      \\ \hline
Optimizer       & Adam & Adam & Adam                         \\
Initial LR      & 0.0001 & 0.0001  & 0.0001                        \\
LR scheduler    & ReduceLROnPlateau        & ReduceLROnPlateau    &-                \\
Loss function   & Cross Entropy     & Cross Entropy & Binary Cross Entropy    \\
Label Smoothing & -     & 0.1    & -  \\
Dropout rate    & 0.5     & -   &0.3 \\
Batch size      & 8      & 8     & 8                        \\
Epoch           & 100     & 200    & 100                     \\
Split ratio     & 70:10:20   & 70:10:20   & 60:15:25                   \\
\hline
\end{tabular}
\end{table}

\subsubsection{Model Selection and Evaluation}

Throughout the training process, model performance was continuously evaluated on the validation set after each epoch. The set of network weights that achieved the highest validation accuracy was saved. This saved checkpoint represents the model at its peak performance on unseen validation data. It was then used for the final unbiased evaluation on the held-out test set. This standard practice ensures that the reported performance metrics are a true reflection of the model's ability to generalize to new data and are not influenced by overfitting to the training set. Each model was trained and evaluated in 5 independent runs and the average performance across 5 runs was recorded. This result is a more stable and realistic estimate of how the model will perform in general.

A repeated-measures Analysis of Variance (ANOVA) followed by Benjamini–Hochberg False Discovery Rate (FDR) corrected post-hoc paired t-tests was conducted to show statistical significance of performance differences among the MRI-based models ($p < 0.05$) and the US-based models ($p < 0.05$). The model with the highest mean accuracy and AUC was selected as the backbone feature extractor for the multimodal model. 

\subsection{Final Comparative Evaluation Protocol}

To ensure a fair and direct comparison, a standardized evaluation protocol was implemented. Three distinct models were evaluated on the same held-out multimodal test set of 40 paired samples. This approach allows comparison between the best-performing standalone unimodal models (trained on larger datasets) and the final fusion model, which leverages the pretraining and is fine-tuned on the smaller paired dataset. The three models compared are as follows:

\begin{enumerate}
    \item \textbf{Unimodal MRI}: This is the top-performing MRI-based 3D model that was fully trained on the large unimodal MRI dataset (n = 793). Its performance on the 40 test MRIs serves as the baseline for what is achievable with MRI alone.
    \item \textbf{Unimodal US}: This is the top-performing US-based 2D model that was trained on the large,  687-sample unimodal US dataset. Its performance on the 40 test US images serves as the baseline for US alone.
    \item \textbf{Multimodal Fusion Model}: This model employed the architectures from the best-performing unimodal experiments as feature extraction backbones. Their weights were initialized from the respective unimodal training runs on the larger datasets. These weights were left unfrozen and the entire fusion model was trained and validated end-to-end on the 120 paired samples. This allowed both branches to be further fine-tuned for extracting complementary features optimized for multimodal fusion. The final performance of the model was evaluated on the 40 paired sample test set.
\end{enumerate}

\subsection{Performance Evaluation Metrics}

The diagnostic performance of the final selected models on their respective test sets was evaluated using a comprehensive set of standard metrics for binary classification tasks. To conduct the qualitative analysis of model errors, a confusion matrix was created in each of the experiments.The confusion matrix gives a clear account of the classification results, which tabulates the true positives (TP), true negatives (TN), false positives (FP), and false negatives (FN). The key performance indicators were:

\begin{itemize}
        \item \textbf{Accuracy: }The proportion of total predictions that were correct.
        \[
        Accuracy = \frac{TP + TN}{TP + TN + FP + FN}
        \]
        \item \textbf{Precision (Positive Predictive Value):}
        The proportion of positive predictions that were actually correct. It helps reduce false positives.
        \[
        Precision = \frac{TP}{TP + FP}
        \]
        \item \textbf{Recall (Sensitivity):}
        The proportion of actual positive cases that were correctly identified by the model. It helps to minimize false negatives.
        \[
        Recall = \frac{TP}{TP + FN}
        \]
        \item \textbf{F1-Score:}
        The harmonic mean of precision and recall, providing a balanced measure of both false positives and false negatives.
        \[
        F1\text{-}Score = \frac{2 \times (Precision \times Recall)}{Precision + Recall}
        \]
        \item \textbf{Area Under the Receiver Operating Characteristic Curve (AUC-ROC): } A measure of the model's overall discriminative ability across all possible classification thresholds. AUC-ROC quantifies the overall performance of the model in a range of 0-1 where 1.0 signifies a perfect classifier, while an AUC of 0.5 indicates performance no better than random chance.
    \end{itemize}

\section{Results}

The unimodal and multimodal deep learning models were evaluated on their respective held-out test sets. The performance metrics were statistically analyzed and compared to find if there are significant differences between the multimodal fusion model and the single-modality models.

\subsection{Unimodal MRI Model Performance}

\begin{figure}[htp!]
    \centering    \includegraphics[width=0.6\textwidth]{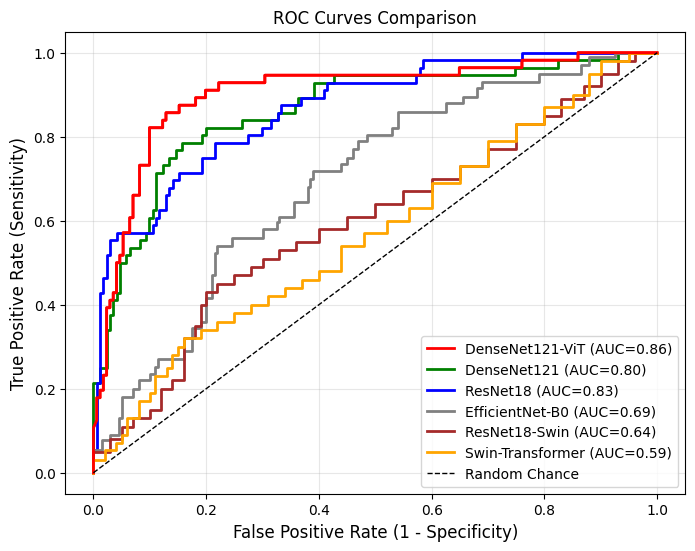} 
    \caption{ROC comparisons of 3D MRI models.~\cite{Ali_2025}}
    \label{fig:mri_roc}
\end{figure}

A set of experiments was conducted to identify the optimal architecture and training configuration for PAS classification using 3D MRI data. Among the various models tested, the Densenet121-ViT hybrid architecture yielded the best overall performance on the hold-out test set of 227 samples. This model achieved an average accuracy of 84.3\% and an AUC of 0.842. The best run for the model had an accuracy of 85\% and an AUC of 0.862. The confusion matrix for the best run showed that it correctly identified 144 out of 171 normal cases and 49 out of 56 PAS cases in the test set. Among the other architectures, the pretrained ResNet18 and DenseNet121 performed well (accuracy $\sim$80\%), while the rest resulted in lower performance, with test accuracies ranging from 62\% to 70\%. Table \ref{tab:model_results_mri} presents the summary of performances from each MRI model. Figure \ref{fig:mri_roc} shows the Receiver Operating Characteristic (ROC) curves for the best run of each model, highlighting differences in sensitivity and specificity across models.

To evaluate the significance of the observed performance differences, repeated measures ANOVA across the six models, followed by post-hoc paired t-tests with FDR correction was conducted. It showed that DenseNet121-ViT outperformed all other models $(p < 0.05)$. DenseNet121 and ResNet18 also performed significantly better than EfficientNet-B0, ResNet18-Swin, and Swin-Transformer, while no significant differences were found between DenseNet121 vs ResNet18 and ResNet18-Swin vs Swin-Transformer $(p > 0.05)$.

\begin{table}[htbp]
\centering
\caption{Performance metrics of unimodal 3D MRI model architectures. Best results in bold.~\cite{Ali_2025}}
\label{tab:model_results_mri}
\renewcommand{\arraystretch}{1.1}
\setlength{\tabcolsep}{5pt}
\scriptsize
\begin{tabular}{lccccc}
\hline
\textbf{Model} & \textbf{Accuracy (\%)} & \textbf{AUC} & \textbf{Precision} & \textbf{Recall} & \textbf{F1-Score} \\
\hline
\pmb{DenseNet121-ViT} & \pmb{85.0 ($84.3 \pm 1.3$)} & \pmb{0.862 ($0.842 \pm 0.012$)} & \pmb{0.799 ($0.790 \pm 0.013$)} & \pmb{0.859 ($0.842 \pm 0.013$)} & \pmb{0.818 ($0.808 \pm 0.014$)} \\
DenseNet121 & 82.8 ($79.5 \pm 2.0$) & 0.804 ($0.766 \pm 0.026$) & 0.770 ($0.732 \pm 0.023$) & 0.802 ($0.764 \pm 0.025$) & 0.783 ($0.743 \pm 0.024$) \\
ResNet18 & 80.2 ($79.3 \pm 1.3$) & 0.829 ($0.783 \pm 0.028$) & 0.759 ($0.738 \pm 0.018$) & 0.832 ($0.781 \pm 0.030$) & 0.772 ($0.750 \pm 0.018$) \\
ResNet18-Swin & 71.3 ($70.0 \pm 1.9$) & 0.644 ($0.600 \pm 0.029$) & 0.629 ($0.599 \pm 0.028$) & 0.642 ($0.604 \pm 0.031$) & 0.634 ($0.601 \pm 0.029$) \\
Swin-Transformer & 72.7 ($69.0 \pm 2.8$) & 0.592 ($0.548 \pm 0.029$) & 0.608 ($0.552 \pm 0.034$) & 0.585 ($0.546 \pm 0.028$) & 0.591 ($0.548 \pm 0.030$) \\
EfficientNet-B0 & 64.8 ($62.8 \pm 1.7$) & 0.692 ($0.604 \pm 0.047$) & 0.596 ($0.573 \pm 0.018$) & 0.622 ($0.592 \pm 0.023$) & 0.593 ($0.569 \pm 0.019$) \\
\hline
\multicolumn{6}{l}{\textit{Note:} Values are reported as \textit{Best (Mean ± Standard Deviation)} over five independent runs.}
\end{tabular}
\end{table}

\subsection{Unimodal US Model Performance}

\begin{figure}[htp!]
    \centering
    \includegraphics[width=0.6\textwidth]{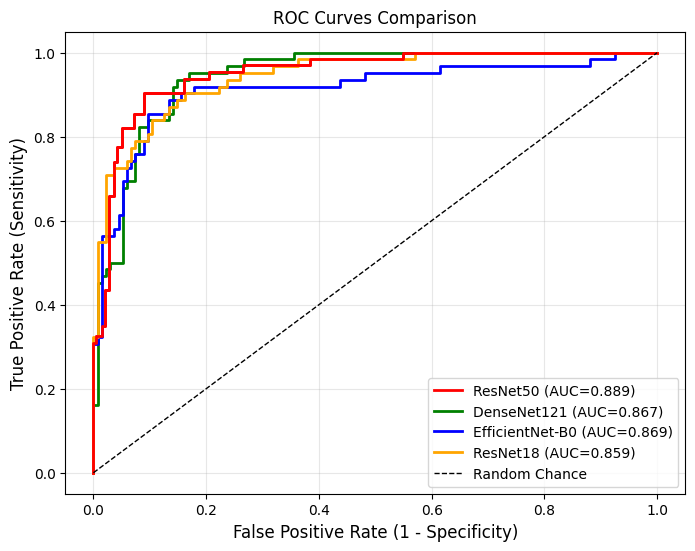} 
    \caption{ROC comparisons of 2D US models.}
    \label{fig:us_roc}
\end{figure}

For the 2D US image classification, several established CNN architectures were evaluated. The ResNet50 model emerged as the best performing, achieving an average accuracy of 86.9\% and an AUC of 0.856 with the best run accuracy of 89.3\% and an AUC of 0.889. The confusion matrix revealed that on the best run, this model correctly classified 123 of the 135 normal cases and 53 of the 62 PAS cases in the test set. Other models, including ResNet18 and EfficientNet, also showed strong performance, with accuracies reaching up to 88\%. Table \ref{tab:tab:model_results_US} presents the summary of the performances from each US model. Figure \ref{fig:us_roc} presents the ROC curves of the best-performing runs for each US model.

\begin{table}[h!]
\centering
\caption{Performance metrics of unimodal 2D US model architectures. Best results in bold.}
\label{tab:tab:model_results_US}
\scriptsize
\begin{tabular}{lccccc}
\hline
\textbf{Model} & \textbf{Accuracy (\%)} & \textbf{AUC} & \textbf{Precision} & \textbf{Recall} & \textbf{F1-Score} \\
\hline
\pmb{ResNet50}      & \pmb{89.3 ($86.9 \pm 1.7$)} & \pmb{0.889 ($0.856 \pm 0.022$)} & \pmb{0.874 ($0.847 \pm 0.020$)} & \pmb{0.883 ($0.855 \pm 0.021$)} & \pmb{0.878 ($0.850 \pm 0.019$)} \\
EfficientNet-B0  & 88.3 ($85.9 \pm 2.2$) & 0.869 ($0.839 \pm 0.027$) & 0.862 ($0.836 \pm 0.025$)& 0.871 ($0.837 \pm 0.029$)& 0.866 ($0.836 \pm 0.027$)\\
ResNet18      & 87.3 ($85.1 \pm 1.4$) & 0.859 ($0.821 \pm 0.028$) & 0.854 ($0.831 \pm 0.015$) & 0.851 ($0.819 \pm 0.022$)& 0.852 ($0.824 \pm 0.018$)\\
DenseNet121   & 84.2 ($82.7 \pm 1.2$)& 0.867 ($0.825 \pm 0.024$)& 0.825 ($0.801 \pm 0.015$)& 0.872 ($0.826 \pm 0.024$)& 0.833 ($0.809 \pm 0.015$)\\
\hline
\multicolumn{6}{l}{\textit{Note:} Values are reported as \textit{Best (Mean ± Standard Deviation)} over five independent runs.}
\end{tabular}
\end{table}

For the US modality, the ANOVA test indicated no statistically significant differences among the models ($p > 0.05$). However, ResNet50 achieved the highest mean Accuracy (86.9\%) and AUC (0.889) and was therefore chosen as the US backbone for multimodal model. Although EfficientNet-B0 showed comparable results, its performance was slightly lower across all metrics.

\subsection{Multimodal Fusion Model Performance}

\begin{figure}[htp!]
    \centering   \includegraphics[width=\textwidth]{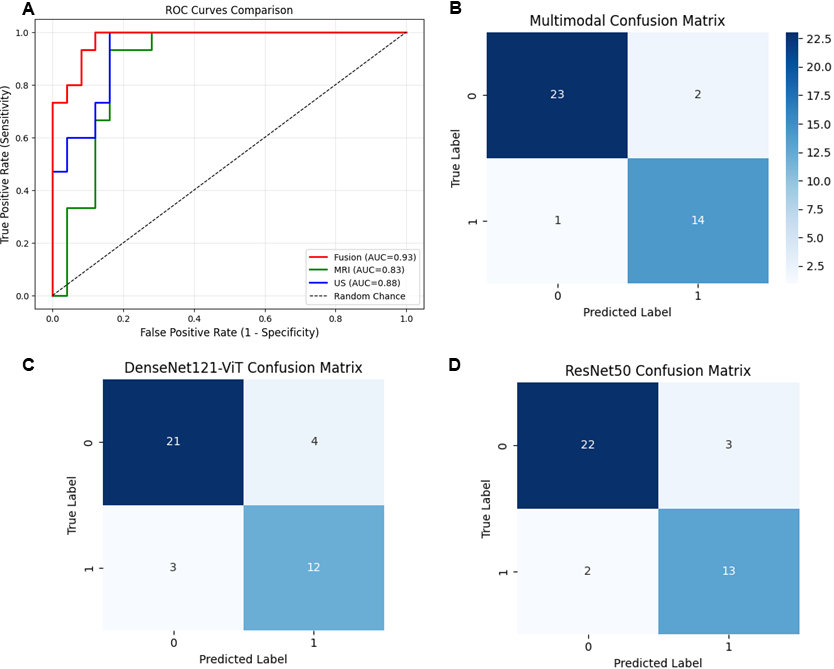} 
    \caption{AUC-ROC and confusion matrix (CM) comparison:  (A) ROC of multimodal and unimodal models, (B) Multimodal CM, (C) MRI-only model CM, (D) US-only CM.}
    \label{fig:ROCCM}
\end{figure}

To thoroughly assess the advantage of data fusion, a direct comparison was established by evaluating the best unimodal models and the fusion model on the exact same unseen test set of 40 paired samples. Overall, the model trained on both MRI and US data outperformed its unimodal counterparts in both accuracy and AUC (Table \ref{tab:fusion_performance}). The feature fusion approach achieved a significantly higher accuracy of 92.5\% and an AUC of 0.927 on its best run. The five run average produced an accuracy of 90.5±1.9\% and AUC of 0.902±0.024. This superior performance was consistent across all metrics, with a precision of 0.892, a recall of 0.903, and an F1-score of 0.897. In the best run, this model correctly identified 23 of 25 normal cases and 14 of 15 PAS cases, resulting in only three overall misclassifications. Figure \ref{fig:ROCCM} presents the best run AUC-ROC comparison and their confusion matrix.

\begin{table}[h!]
\centering
\caption{Comparison of Unimodal and Multimodal Model Performance on same test set.}
\label{tab:fusion_performance}
\scriptsize
\begin{tabular}{lccccc}
\hline
\textbf{Model}          & \textbf{Accuracy (\%)} & \textbf{AUC} & \textbf{Precision} & \textbf{Recall} & \textbf{F1-Score} \\ \hline
Unimodal MRI      & 82.5 ($80.5 \pm  1.9$)   & 0.825 ($0.800 \pm  0.018$)       & 0.813 ($0.793 \pm  0.020$)            & 0.820 ($0.799 \pm  0.018$)           & 0.816 ($0.795 \pm  0.019$)            \\
Unimodal US       & 87.5 ($86.0 \pm  1.2$)          & 0.879 ($0.844 \pm  0.024$)       & 0.865 ($0.859 \pm  0.012$)              & 0.873 ($0.843 \pm  0.022$)          & 0.868 ($0.847 \pm  0.016$)            \\
\textbf{Multimodal} & \pmb{92.5 ($90.5 \pm  1.9$)}         & \pmb{0.927 ($0.902 \pm  0.024$)}       & \pmb{0.917 ($0.892 \pm  0.022$)}             & \pmb{0.927 ($0.903 \pm  0.022$)}          & \pmb{0.921 ($0.897 \pm  0.022$)}            \\ \hline
\multicolumn{6}{l}{\textit{Note:} Values are reported as \textit{Best (Mean ± Standard Deviation)} over five independent runs.}
\end{tabular}
\end{table}

\subsection{Statistical Analysis}

To compare model performance across evaluation metrics, a repeated-measures ANOVA was conducted for each metric. When significant differences were found, post-hoc paired t-tests were performed with FDR correction using Benjamini–Hochberg method to control for multiple comparisons. Table \ref{tab:anova_fdr} summarizes the statistical results across all metrics.

\begin{table}[ht]
\centering
\caption{Statistical comparison of model performance across five evaluation metrics using FDR-corrected post-hoc pairwise comparisons. Significant p-values ($p < 0.05$) are shown in bold.}
\small
\begin{tabular}{lccc}
\hline
\textbf{Metric} &  \textbf{Fusion vs MRI} & \textbf{Fusion vs US} & \textbf{MRI vs US} \\
\hline
Accuracy   & \pmb{$<0.001$} & \textbf{0.026} & \textbf{0.013} \\
AUC        & \pmb{$<0.001$} & \textbf{0.021} & \textbf{0.030} \\
Precision & \textbf{0.003}  & 0.270& \textbf{0.023} \\
Recall     & \pmb{$<0.001$} & \textbf{0.006} & \textbf{0.030}\\
F1         & \textbf{0.002} & \textbf{0.036} & \textbf{0.013} \\
\hline
\end{tabular}
\label{tab:anova_fdr}
\end{table}

The ANOVA revealed significant differences among the three models for each of the metrics. FDR-corrected post-hoc analyses showed that the fusion model significantly outperformed both the MRI-only and US-only models in terms of Accuracy, AUC, Recall, and F1-score (all $p < 0.05$). For Precision, the fusion model performed significantly better than the MRI (p = 0.003), while the difference between fusion and US models was not significant (p = 0.270). 

\subsection{Interpretability Analysis}

\begin{figure}[htp!]
    \centering
    \includegraphics[width=0.8\textwidth]{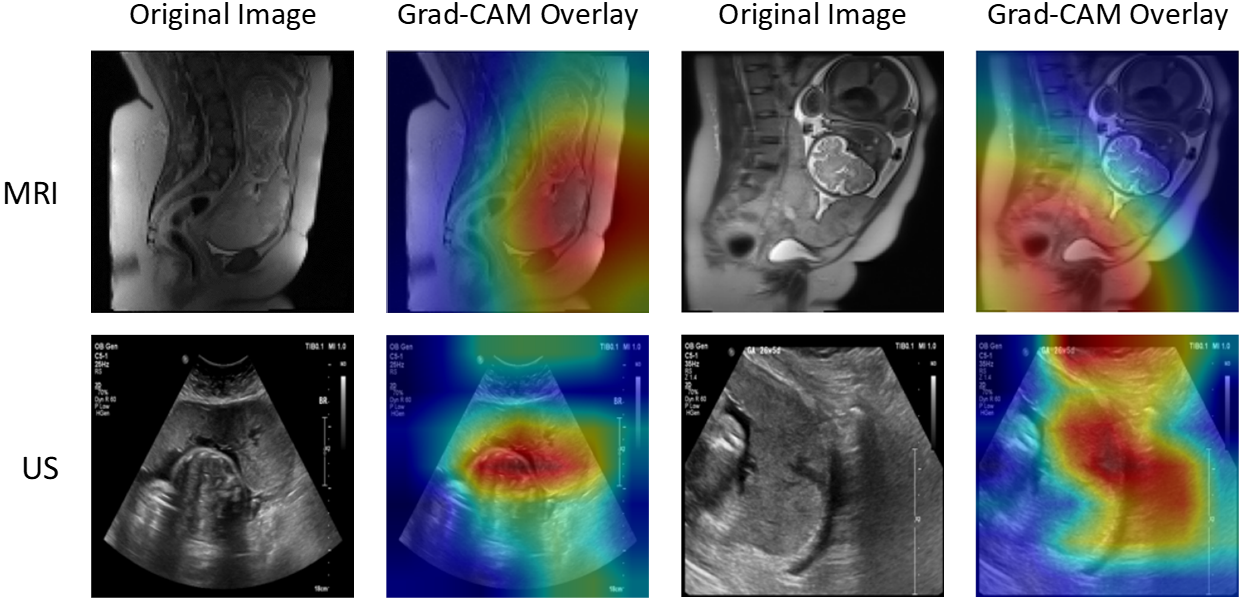} 
    \caption{Grad-CAM heatmaps for MRI and US feature extractors.}
    \label{fig:grad_cam}
\end{figure}

To evaluate the interpretability of the proposed model, Grad-CAM~\cite{selvaraju_2020} visualizations were generated for representative MRI and US cases. These visualizations highlight the regions in the input images that contributed the most to the model's decision. Figure~\ref{fig:grad_cam} illustrates the MRI and US feature extractors' attention maps alongside the original images for qualitative assessment.

The Grad-CAM was computed from the last 3D convolutional layer of the MRI and US feature extractors. The resulting 3D activation maps were visualized as 2D slice overlays to highlight anatomical regions most contributing to the classification. Each Grad-CAM output was resized, normalized and superimposed on the corresponding input image to form a heatmap overlay. For the US modality, the visualization was displayed side-by-side with the original image. For MRI volumes, Grad-CAM was computed across slices, and representative slices from different depths were visualized.

By visualizing the Grad-CAM heatmaps for MRI and US images, it was observed that the multimodal model primarily focused on anatomically relevant regions associated with PAS~\cite{Romeo_2021}. In MRI, the attention was concentrated on the uterine wall and areas of abnormal placental invasion, while in US, the model highlighted the lower uterine segment and the interface between the placenta and myometrium, which are known key signs of PAS~\cite{Romeo_2021, Arakaza_2023}. These findings suggest that the model correctly identifies the critical regions that are indicative of PAS. The attention maps provide insights into the decision-making process of the model, supporting the reliability of the predictions and enhancing interpretability, which is crucial for clinical application.

\section{Discussion}

\subsection{Unimodal and Multimodal Performance Analysis}

The unimodal experiments revealed a notable outcome, the stronger performance of the 2D US-based ResNet50 model (89.3\%, AUC 0.889) relative to the more complex 3D MRI-based hybrid DenseNet121-ViT model (85\%, AUC 0.862) on their own unimodal test sets. MRI is often considered a more reliable imaging modality for assessing deep myometrial invasion. The performance gap is likely influenced less by the diagnostic capability of the modalities and more by the capacity of the learning methods applied. 2D CNNs like ResNet50 are highly optimized and benefit greatly from transfer learning on huge 2D medical image databases like RadImageNet, making them powerful feature extractors that are less likely to overfit~\cite{Mei_2022}. In contrast, 3D models are well known to require substantially larger datasets, and the complex 3D DenseNet-ViT architecture may have been under trained given the available data. This suggests that for PAS detection, within the context of this study, the relevant diagnostic features may be more readily identifiable from 2D US data, at which fine-tuned 2D CNNs excel. This promising performance of the unimodal US model therefore warrants further investigation to rigorously assess its potential as a first-line screening tool. 

Building on the unimodal baselines, this study demonstrates the significant benefit of a multimodal deep learning framework for the accurate and non-invasive diagnosis of PAS. To accurately assess the added value of multimodal fusion, a direct evaluation of the multimodal model and the unimodal models on a same test set was essential. The true strength of this study lies in the multimodal model's ability to surpass these unimodal baselines. Evaluation on the dedicated paired test set (n = 40) showed that while unimodal models using either MRI or US data can achieve strong performance, the integration of both modalities through an intermediate fusion architecture led to superior diagnostic performance. The multimodal model results indicates that MRI and US offer complementary diagnostic information by achieving an accuracy of 92.5\% and an AUC of 0.927 on the test set, outperforming the unimodal MRI model (accuracy 82.5\%, AUC 0.825) and US model (accuracy 87.5\%, AUC 0.879).

Statistical analysis demonstrates that integrating MRI and US features gives a statistically significant performance improvement over unimodal models. The multimodal approach consistently achieved higher Accuracy, AUC, Recall, and F1-score, indicating improved overall detection capability and sensitivity. The lack of significant improvement in Precision compared to the US-only model suggests that while fusion helps identify more true positives, it may not greatly reduce false positives. Overall, these findings highlight the complementary nature of MRI and US modalities and support the effectiveness of multimodal fusion for robust classification. This supports the primary hypothesis of the study that combining imaging modalities could provide more comprehensive and accurate diagnostic information than either modality alone.

\subsection{Comparison with Prior Work}

Table \ref{tab:pas_comparison} compares the unimodal baseline results with related studies on PAS detection. The reported results are obtained using internal independent test sets to provide a comparative analysis. Studies are grouped by imaging modality and the table highlights key methodological differences. Existing MRI-based deep learning studies demonstrated promising performance with accuracy ranging from 75-84.3\%. But they are limited by factors such as small datasets, class imbalance~\cite{Peng_2023}, slice-level data splits~\cite{Xu_2022}, or multi-stage pipelines involving manual annotation and extensive preprocessing~\cite{Wang_2023}, which can introduce data leakage, class imbalance bias, and limited scalability. Whereas, the unimodal MRI in this study used end-to-end hybrid DenseNet121–ViT architecture directly on whole 3D MRI volumes, with patient-level data splitting, balanced training and evaluation on an independent test set that eliminate manual feature engineering and improve model generalizability. With its automated end-to-end approach, the proposed model achieves competitive performance (85\% accuracy, AUC 0.86) and outperforms the prior studies.

\begin{table*}[t]
\centering
\caption{Comparison of unimodal baseline results with related PAS detection studies.}
\label{tab:pas_comparison}
\renewcommand{\arraystretch}{1.2}
\setlength{\tabcolsep}{6pt}
\scriptsize
\begin{tabular}{m{1.9cm} c m{1cm} l l c c l}
\hline
\textbf{Study} & \textbf{Year} & \textbf{Modality} & \textbf{Dataset} & \textbf{Model} & \textbf{Accuracy} & \textbf{AUC} & \makecell[l]{\textbf{Notes}} \\
\hline

\multicolumn{8}{l}{\textbf{\textit{MRI-based PAS Detection}}} \\
\hline
Peng et al.~\cite{Peng_2023} & 2023 & 3D MRI & 324 (206 PAS) & 3D ResNet & 75.0\% & 0.861 & No class balancing \\\hline
Wang et al.~\cite{Wang_2023} & 2023 & 3D MRI & 540 (170 PAS) &  \makecell[l]{3D nnU-Ne,\\2D DenseNet-PAS} & 84.3\% & 0.860 & \makecell[l]{Manual annotation;\\Extensive preprocessing}\\\hline
Xu et al.~\cite{Xu_2022} & 2022 & 2D MRI & 321 (179 PAS) & 2D ResNet50 & 82.5\% & -- & \makecell[l]{Slice-level split;\\Non-independent test set}\\\hline
\textbf{Unimodal MRI (This Study)} & \textbf{2025} & \textbf{3D MRI }& \textbf{1,133 (280 PAS)} & \textbf{3D DenseNet121-ViT} & \textbf{85.0\%} & \textbf{0.862} & \makecell[l]{\textbf{End-to-end pipeline;}\\\textbf{Patient-level split;}\\\textbf{Balanced training set;}\\\textbf{Independent test set}}\\
\hline\hline
\multicolumn{8}{l}{\textbf{\textit{US-based PAS Detection}}} \\
\hline 
Young et al.~\cite{Young_2024} & 2024 & \makecell[l]{2D US\\Features} & 154 (77 PAS) & \makecell[l]{Extra Trees \\ Linear Classifier} & \makecell{83.0\% \\ 88.7\%} & \makecell{0.90 \\ --} & \makecell[l]{Manual ROI extraction;\\Feature-based model}\\\hline
\textbf{Unimodal US (This Study)} & \textbf{2025} & \textbf{2D US} & \textbf{983 (214 PAS)} & \textbf{2D ResNet50} & \textbf{89.3\%} & \textbf{0.89} & \makecell[l]{\textbf{End-to-end pipeline;}\\\textbf{Automated feature}\\\textbf{extraction}}\\

\hline
\end{tabular}
\end{table*}

For US-based approaches, there is no prior studies with deep learning for PAS detection using US images. Existing US work~\cite{Young_2024} relies on manually extracted ROI from US images and ROI-based machine learning methods on small dataset with results (accuracy 84-92\%) reported on very small test sample (39 images). For this reason, cross-validation performance of accuracy 83-88.7\% is compared as a more stable indicator of model capability. In contrast, the unimodal US model in this study automatically learns diagnostic patterns from the full 2D US image on a larger dataset, achieving high performance (89.3\% accuracy, AUC 0.89) with greater generalizability.

Additionally, the proposed multimodal model performance was compared with relevant studies. To the best of current knowledge, there are no prior studies published that perform feature-level multimodal fusion of US and MRI images for PAS detection. The only related multimodal work~\cite{Ye_2022} integrates 2D MRI images with MRI-derived radiomic features and clinical variables using decision-level fusion and ResNet34 encoder with machine learning to get 85.2\% accuracy and 0.857 AUC. In comparison, the proposed end-to-end multimodal model uses feature-level fusion with deep learning architectures that take both 3D MRI and 2D US as inputs to achieve 92.5\% accuracy and AUC of 0.927. This study differs fundamentally from image-based multimodal learning and comparison is intended to highlight methodological differences rather than performance. This result also holds up well when compared with multimodal fusion studies in other medical areas, such as breast cancer (AUC 0.898)~\cite{Holste_2021} and Alzheimer's disease~\cite{Venugopalan_2021}. The intermediate (or feature-level) fusion architecture was chosen deliberately as it is a technique that has proven effective in various medical imaging applications~\cite{Holste_2021, Pei_2023}. This strategy allows the model to first learn high-level, abstract features from each modality independently before combining them to identify complex inter-modal relationships.

\section{Conclusion}

In conclusion, this study successfully developed and validated a multimodal deep learning model for the diagnosis of PAS that integrates information from both MRI and US images. The results suggest that the proposed intermediate fusion architecture has the potential to outperform models trained on either modality alone, achieving a diagnostic accuracy of 92.5\% and an AUC of 0.927 on a test set. These results highlight the complementary value of combining the anatomical detail of MRI with US functional information. With further validation, this framework has the potential to become a powerful decision support tool in clinical practice, improving the precision of prenatal PAS diagnosis and ultimately contributing to improved maternal and neonatal outcomes.

While the study showed promising results, several limitations are recognized. Its retrospective design based on data from a single institution may limit the generalizability of the proposed trained models. Future work is needed to evaluate the performance on external data from other sites with different imaging protocols. Additionally, the size of the paired multimodal dataset was modest, a common challenge in medical AI. Hence, validation on larger, multi-center datasets is needed to confirm these preliminary findings. Additionally, expanding the multimodal approach to include clinical data, which has shown promise in other studies, would also be a valuable addition.

\section*{Acknowledgments}

The data collected for this study was approved by the Research Ethics Committee at King Abdulaziz University (Reference No. HA-02-J-008 ) on September 11, 2023, ensuring compliance with ethical guidelines for participant rights and confidentiality.

\bibliographystyle{unsrt}
\bibliography{references}

\end{document}